# Dynamic interactions in terms of senders, hubs, and receivers (SHR) using the singular value decomposition of time series: Theory and brain connectivity applications

Roberto D. Pascual-Marqui[1,2] and Rolando J. Biscay-Lirio[3,4]
1: The KEY Institute for Brain-Mind Research, University Hospital of Psychiatry, Zurich, Switzerland
2: Department of Neuropsychiatry, Kansai Medical University, Osaka, Japan
3: Institute for Cybernetics, Mathematics, and Physics, Havana, Cuba
4: DEUV-CIMFAV, Facultad de Ciencias, Universidad de Valparaiso, Chile

Corresponding author:
Roberto D. Pascual-Marqui
The KEY Institute for Brain-Mind Research
University Hospital of Psychiatry
Lenggstrasse 31
CH-8032, Zurich
Switzerland
pascualm@key.uzh.ch

**Abstract**: Understanding of normal and pathological brain function requires the identification and localization of functional connections between specialized regions. The availability of high time resolution signals of electric neuronal activity at several regions offers information for quantifying the connections in terms of information flow. When the signals cover the whole cortex, the number of connections is very large, making visualization and interpretation very difficult. We introduce here the singular value decomposition of time-lagged multiple signals, which localizes the senders, hubs, and receivers (SHR) of information transmission. Unlike methods that operate on large connectivity matrices, such as correlation thresholding and graph-theoretic analyses, this method operates on the multiple time series directly, providing 3D brain images that assign a score to each location in terms of its sending, relaying, and receiving capacity. The scope of the method is general and encompasses other applications outside the field of brain connectivity.

### Introduction

Electric neuronal activity can be recorded invasively from the human brain by means of intracranial electrodes (see e.g. Crone et al, 2009). The time series of electric potentials provided by such electrodes can have very high time resolution. In particular, if a pair of such electrodes is placed with a small separation distance, then the local electric potential difference is a gradient, and is proportional to the current density vector projected onto the line joining the electrodes, according to Ohm's law (see e.g. Sarvas, 1987):

**Eq. 1** $\quad -\sigma \nabla \varphi = \mathbf{J}$

where the electric neuronal activity is given by the current density $\mathbf{J}$, $\sigma$ is conductivity, $\varphi$ is electric potential, and $\nabla$ is the gradient operator.

Such time series contain local information on brain function.

In practice, it is very desirable to be able to obtain such information non-invasively. This can be achieved by computing the current density in the brain, from non-invasive electric potential differences recorded on the scalp, i.e. from the EEG.

In particular, the method of choice in this paper for solving this inverse problem is exact low resolution electromagnetic tomography (eLORETA), see e.g. Pascual-Marqui (2007, 2009). This method is a multivariate linear solution to the EEG inverse problem, and is endowed with the property of exact localization response when probed with Dirac deltas (which in this case are test dipoles located anywhere in the brain). Due to the principles of linearity and superposition, the method will perform well with any distribution of current density, albeit with low spatial resolution.

Regardless of the technique used for obtaining the current density, either invasive (with intracranial electrodes) or non-invasive (computed from the EEG by means of a validated inverse solution), the method to be described here can be used for revealing connections in the brain. In particular, the new method is capable of localizing and distinguishing senders, hubs, and receivers of information transmission in the brain.

**Stationary case**

Let $U_{i,t}$ denote the current density at the *i*-th voxel, at time $t$; with $i=1...N_V$, $N_V$ denoting the number of cortical voxels; $t=1...N_T$, $N_T$ denoting the number of time frames (discrete time samples). At the *i*-th voxel, consider the univariate autoregressive model:

Eq. 2 $$U_{i,t} = \sum_{k=1}^{Q} a_{i,k} U_{i,t-k} + V_{i,t}$$

where $Q \geq 2$ denotes the global autoregressive order, $a_{i,k}$ are the auto-regression coefficients, and $V_{i,t}$ is the innovation time series.

In a first step, it will be required to fit the model in Eq. 2 separately to each voxel, in order to obtain the innovations. The vector of innovation time series containing all voxels is denoted as:

Eq. 3 $\quad \mathbf{V}_t \in \mathbb{R}^{N_V \times 1}$

defined for $t = (Q+1)...N_T$.

Now define the vector:

Eq. 4 $$\mathbf{Z}_t = \begin{pmatrix} \mathbf{V}_t \\ \mathbf{V}_{t-1} \\ . \\ . \\ \mathbf{V}_{t-Q} \end{pmatrix} \in \mathbb{R}^{[(Q+1)N_V] \times 1}$$

defined for $t = (2Q+1)...N_T$, which contains the innovation current density at time $t$ and its past, back to $Q$ time units into the past.

Next, form the data matrix:

Eq. 5 $\quad \mathbf{Z} = \begin{pmatrix} \mathbf{Z}_{2Q+1} & \mathbf{Z}_{Q+2} & ... & \mathbf{Z}_{N_T} \end{pmatrix} \in \mathbb{R}^{[(Q+1)N_V] \times (N_T - 2Q)}$

and normalize each row, i.e. the time series at each row of the data matrix $\mathbf{Z}$ should have zero mean and unit variance. When $\mathbf{Z}$ is normalized in this way, then $\mathbf{Z}\mathbf{Z}^T$ corresponds to the common cross-correlation matrix for multivariate time series.

Consider the singular value decomposition (SVD) of $\mathbf{Z}$:

**Eq. 6** $\quad \mathbf{Z} = \mathbf{L}\Lambda\mathbf{R}^T$

with $\mathbf{L} \in \mathbb{R}^{[(Q+1)N_V] \times K}$ containing the left eigenvectors, $\mathbf{R} \in \mathbb{R}^{(N_T-Q) \times K}$ containing the right eigenvectors, $\Lambda \in \mathbb{R}^{K \times K}$ is diagonal and contains the eigenvalues in descending order, and $K = \min\{[(Q+1)N_V], (N_T-2Q)\}$. Both $\mathbf{L}$ and $\mathbf{R}$ are orthonormal, i.e. $\mathbf{L}^T\mathbf{L} = \mathbf{I}$ and $\mathbf{R}^T\mathbf{R} = \mathbf{I}$.

The main feature of interest is the first column of $\mathbf{L}$, i.e. the first left eigenvector denoted as $\Gamma \in \mathbb{R}^{[(Q+1)N_V] \times 1}$, corresponding to the largest eigenvalue. As can be seen from the structure of the data matrix defined by Eq. 4, we note that the first $N_V$ elements of $\Gamma$ correspond to the present, the next $N_V$ elements correspond to one step in the past, and so on:

**Eq. 7** $\quad \Gamma = \begin{pmatrix} \Gamma_t \\ \Gamma_{t-1} \\ . \\ . \\ . \\ \Gamma_{t-Q} \end{pmatrix} \in \mathbb{R}^{[(Q+1)N_V] \times 1}$

Definitions:

1. The elements of $\Gamma_t \in \mathbb{R}^{N_V \times 1}$ quantify the "receiving" function at each voxel.
2. The elements of $\Gamma_{t-Q} \in \mathbb{R}^{N_V \times 1}$ quantify the "sending" function at each voxel.
3. The elements of $\Gamma_{t-k} \in \mathbb{R}^{N_V \times 1}$, for $k = 1...(Q-1)$, quantify the "hub" function at each voxel.

The first right eigenvector of $\mathbf{R}$, denoted as $\mathbf{R}_1 \in \mathbb{R}^{(N_T-2Q) \times 1}$, is the time series that expresses the dynamics of the senders, hubs, and receivers given by $\Gamma$.

**Notes and motivation**

At the heart of the concepts of "senders, hubs, and receivers" (SHR), rests Granger causality (1967), which is based on the analysis of the innovation time series, and not the original time series. This is one reason for computing the SVD on the innovation time series.

Furthermore, the explicit dependence of the current density at a given voxel with its own past might be a confounding factor for the hub function: a voxel sending and receiving large amounts of information is by definition a hub, even if it sends and receives from its own self. Therefore, by partialling out the univariate auto-dependence at each voxel, the residuals should be better at characterizing the hub.

At the heart of using the largest left eigenvector of the data matrix constructed in Eq. 5 is the method advocated by Worsley et al (2005). In that paper, it is shown that this eigenvector reveals functional connectivity between voxels. By augmenting the time series at each voxel with their time-shifted pasts, we now additionally have information on "Granger-causal functional connectivity".

The structure of the data matrix as defined in Eq. 4 carries over to the left eigenvector. This allows the interpretation of its different blocks in terms of senders, hubs, and receivers. For instance, note that:

1. $\Gamma_t$ is the ultimate receiver, (Granger-) causally being influenced by the pasts $\Gamma_{t-1}...\Gamma_{t-Q}$.

2. $\Gamma_{t-Q}$ is the ultimate sender, (Granger-) causally influencing all the immediate future $\Gamma_t...\Gamma_{t-Q+1}$.

3. The intermediate time blocks $\Gamma_{t-1}...\Gamma_{t-Q+1}$ can send and receive, and are therefore related to hubs, i.e. to relay stations.

If the global auto-regressive order $Q$ is set to 1, then the hub function is not defined, although senders and receivers remain well defined.

In practice, when the number of voxels is very large compared to the number of time samples, it might be efficient to compute only the largest left and right eigenvectors by applying the power method to the matrix $\mathbf{Z}$.

In the individual univariate time series for each voxel (Eq. 2), it is possible to consider the use of different autoregressive order values, as long as they are at least as large as the global $Q$.

**Locally stationary case**

It will be assumed that the current density time series can be repeatedly sampled, and that each sample starts at a repeated event. For instance, each sample might correspond to the presentation of a visual stimulus, and all samples are time locked to the moment of stimulus onset.

Now let $U_{i,t,j}$ denote the current density at the *i*-th voxel, at time $t$, for the *j*-th sample (e.g. for the *j*-th stimulus presentation), with $j=1...N_S$, $N_S$ denoting the number of samples (e.g. the number of stimuli).

We will consider local univariate autoregressive models of order $Q$, specified at the target time $\tau$, defined at values $\tau=(2Q+1)...N_T$. The model will be locally valid for time instants in the immediate past of the target time, for $(\tau-2Q)\leq t \leq \tau$.

In what follows, the target time $\tau$ is considered fixed.

For the *i*-th voxel, the model is:

**Eq. 8** $$U_{i,t,j} = \sum_{k=1}^{Q} a_{\tau,i,k} U_{i,t-k,j} + V_{\tau,i,t,j}, \text{ for } (\tau-2Q)\leq t \leq \tau$$

where $a_{\tau,i,k}$ are the auto-regression coefficients, and $V_{\tau,i,t,j}$ is the innovation time series.

Note that the available data for estimating this model consists of all the samples $j=1...N_S$, but only the local time instants $(\tau-2Q)\leq t \leq \tau$. From here we estimate the coefficients $a_{\tau,i,k}$, and the innovations for all samples ( $j=1...N_S$ ), at $t=\tau$, $t=\tau-1$, ... ,$t=\tau-Q$. The vector of innovation time series containing all voxels is denoted as:

**Eq. 9** $\quad \mathbf{V}_{\tau,t,j} \in \mathbb{R}^{N_V \times 1}$

defined for local times $\tau - Q \leq t \leq \tau$, and for all samples $j = 1...N_S$.

Now define the vector:

**Eq. 10** $\quad \mathbf{Z}_{\tau,j} = \begin{pmatrix} \mathbf{V}_{\tau,t,j} \\ \mathbf{V}_{\tau,t-1,j} \\ . \\ . \\ \mathbf{V}_{\tau,t-q,j} \end{pmatrix} \in \mathbb{R}^{[(Q+1)N_V] \times 1}$

which contains the innovation current density for the *j*-th sample, at time $t$ and its past, back to $Q$ time units into the past.

Next, form the data matrix:

**Eq. 11** $\quad \mathbf{Z}_\tau = \begin{pmatrix} \mathbf{Z}_{\tau,1} & \mathbf{Z}_{\tau,2} & ... & \mathbf{Z}_{\tau,N_S} \end{pmatrix} \in \mathbb{R}^{[(Q+1)N_V] \times N_S}$

and normalize each row, i.e. the sample values at each row of the data matrix $\mathbf{Z}_\tau$ should have zero mean and unit variance. When $\mathbf{Z}_\tau$ is normalized in this way, then $\mathbf{Z}_\tau \mathbf{Z}_\tau^T$ corresponds to the cross-correlation matrix at target time $\tau$ for local multivariate time series.

As in the stationary case above, the first left eigenvector of the SVD of $\mathbf{Z}_\tau$, now denoted as $\mathbf{\Gamma}_\tau$, contains all the relevant information on senders, hubs, and receivers, at each target time $\tau$.